\documentclass[showpacs,preprintnumbers, nofootinbib]{revtex4}
\usepackage{amssymb}

\usepackage{fontenc}
\usepackage{graphicx}
\usepackage[dvips]{hyperref}
\usepackage{amsmath}

\begin{document}

\title{Quasinormal modes and stability analysis for $z=4$ Topological black hole in $4+1$ dimensional Horava-Lifshitz gravity}
\author{Ram\'{o}n Becar}
\email{rbecar@uct.cl}
\affiliation{Departamento de Ciencias Matem\'{a}ticas y F\'{\i}sicas, Universidad Cat\'{o}%
lica de Temuco, Montt 56, Casilla 15-D, Temuco, Chile.}
\author{P. A. Gonz\'{a}lez}
\email{pgonzalezm@ucentral.cl}
\affiliation{Escuela de Ingenier\'{\i}a Civil en Obras Civiles. Facultad de Ciencias F%
\'{\i}sicas y Matem\'{a}ticas, Universidad Central de Chile, Avenida Santa
Isabel 1186, Santiago, Chile}
\affiliation{Universidad Diego Portales, Casilla 298-V, Santiago, Chile.}
\author{Yerko V\'{a}squez}
\email{yvasquez@ufro.cl}
\affiliation{Departamento de Ciencias F\'{\i}sicas, Facultad de
Ingenier\'{i}a, Ciencias
y Administraci\'{o}n, Universidad de La Frontera, Avenida Francisco Salazar
01145, Casilla 54-D, Temuco, Chile.}

\date{\today}

\begin{abstract}
We study $z=4$ Topological black hole in $4+1$ dimensional Horava-Lifshitz gravity and we calculate analytically the quasinormal modes of scalar perturbations and from these quasinormal modes we show that $z=4$ Topological black hole in $4+1$ dimensional Horava-Lifshitz gravity is stable.
\end{abstract}

\maketitle


\section{Introduction}

Few years ago, P. Horava proposed \cite{Horava:2008jf, Horava:2008ih, Horava:2009uw} a new class of quantum field theory as an UV complete theory of gravity, which is non-relativistic and exhibit anisotropic
scaling between space and time. The Horava theory is power-counting renormalizable. However, the Lorentz symmetry  is broken at short distances. Moreover, the theory can be reduced to Einstein's gravity theory with a cosmological constant in the infrared (IR) limit. A lot of attention has been focused on this gravity theory. The Lifshitz-type black holes may provide a way to generalize $AdS/CFT$ correspondence to non-relativistic condensed matter physics, \cite{Horava:2009uw,Lu:2009em,Cai:2009pe,Kehagias:2009is,Ghodsi:2009zi,Balasubramanian:2009rx,Taylor:2008tg,Matulich:2011ct, Koutsoumbas:2010pt}.
Horava-Lifshitz with dynamical exponent $z=4$ \cite{Horava:2009uw} has acquired a growing interest,  in Ref. \cite {Horava:2009if} was shown using the numerical Causal Dynamical Triangulations (CDT) approach \cite{Ambjorn:2005db} for quantum gravity in $3+1$ dimensions that the definition of "spectral dimension" can be extended to theories on smooth spacetimes with anisotropic scaling. 
Therefore, CDT approach to lattice gravity may in fact be a lattice version of the quantum gravity at $z=4$ Lifshitz point. 

In this paper, we  focus our investigation in the$z=4$ Topological black hole in $4+1$ dimensional Horava-Lifshitz gravity, in the $IR$ region obtained via detailed balance condition by \cite{Cai:2009ar}, these black holes are similar to the topological black hole solution in the five-dimensional Chern-Simons gravity \cite{Cai:1998vy}, or in the five-dimensional Gauss-Bonnet gravity with a special Gauss-Bonnet coefficient \cite{Cai:2001dz}. The lagrangian for this theory is given by \cite{Cai:2009ar} 
\begin{equation}
\mathcal{L}=\mathcal{L}_0+\mathcal{L}_1~,
\end{equation}
where
\begin{equation}
\mathcal{L}_0=\sqrt{g}N\left\{\frac{2}{\kappa^{2}}(K^{ij}K_{ij}-\lambda K^{2})+\frac{
\kappa^{2}\mu^{2}(\Lambda_{W}R-2\Lambda^{2}_W)}{4(1-4\lambda)}\right\}~,
\end{equation}
\begin{equation}
\mathcal{L}_1=-\sqrt{g}N\frac{\kappa^{2}}{8}\left\{\mu^{2}G_{ij}G^{ij}+\frac{2\mu}{M}
G^{ij}L_{ij}+\frac{2\mu}{M}\Lambda_{W}L+\frac{1}{M^{2}}L^{ij}L_{ij}-\tilde{
\lambda}\left(\frac{L^{2}}{M^{2}}-\frac{2\mu L}{M}(R-4\Lambda_{W})+
\mu^{2}R^{2}\right)\right\}~,
\end{equation}
\begin{equation}
L=2(1+3\beta)\nabla^{2}R~,
\end{equation}
the expressions $G^{ij}$ and $L^{ij}$ are given by
\begin{equation}
G^{ij}=R^{ij}-\frac{1}{2}g^{ij}R~,
\end{equation}
\begin{equation}
L^{ij}=(1+2\beta)(g^{ij}\nabla^2-\nabla^i\nabla^j)R+\nabla^2G^{ij}+2\beta R(R^{ij}-\frac{1}{4}g^{ij}R)+2(R^{imjn}-\frac{1}{4}g^{ij}R^{mn})R_{mn}~.
\end{equation}
and the term  $K^{ij}$ is given by
\begin{equation}
K_{ij}=\frac{1}{2N}(\dot{g}_{ij}-\nabla_iN_j-\nabla_jN_i)~,
\end{equation}
for a $D+1$ dimensional metric written in the following form
\begin{equation}
ds^2=-N^2c^2dt^2+g_{ij}(dx^i-N^idt)(dx^j-N^jdt)~,
\end{equation}
where $i=1,..., D$ and c is the speed of light.
Here, we focus our investigation in the $z=4$ Topological black hole in $4+1$ dimensional Horava-Lifshitz gravity, in the $IR$ region, \cite{Cai:2009ar}. Which, for
$\beta=-\frac{1}{3}$, \cite{Cai:2009ar}, is given by the metric
\begin{equation}  \label{metric}
\ ds=-\tilde{N}^{2}f(r)dt^{2}+\frac{dr^{2}}{f(r)}+r^{2}d\Omega^{2}_{k},
\end{equation}
where $f(r)=-\frac{\Lambda_W r^{2}}{3}+k \pm\sqrt{c_{0}}$, $\tilde{N}(r)=1$, $c_0\geq0$ and
$d\Omega^{2}_{k}$ is the three-dimensional Einstein manifold with constant
scalar curvature $6k$, which we may choose to be $k=0$,$\pm1$, corresponding to a spherical, plane or hyperbolic section, respectively. Which, is similar to the topological black hole solution in the five-dimensional Chern-Simons gravity \cite{Cai:1998vy}, or in the five-dimensional Gauss-Bonnet gravity with a special Gauss-Bonnet coefficient \cite{Cai:2001dz}. 

According to the AdS/CFT correspondence~\cite{Maldacena:1997re}, the quasinormal modes (QNMs)  \cite{Regge:1957td, Zerilli:1971wd, Zerilli:1970se, Kokkotas:1999bd, Nollert:1999ji, Berti:2009kk, Konoplya:2011qq}  allow us to obtain the relaxation time of a thermal state of the conformal theory at the boundary, which is
proportional to the inverse of imaginary part of the QNMs of the dual
gravity background ~\cite{Horowitz:1999jd}. In this work we investigate scalar perturbations in the background of $z=4$ Topological black hole in $4+1$ dimensional Horava-Lifshitz gravity, see Refs. \cite{Chan:1999sc, Wang:2001tk, Aros:2002te, Gonzalez:2010vv, Gonzalez:2012de, Gonzalez:2012xc} for scalar perturbations in the background of a Topological black hole, and we present the exact QNMs, considering Dirichlet and Neumann boundary conditions. Then, we study the stability of this black hole analyzing the imaginary part of the QNMs.

The plan of the work is as follows: In section II we calculate the exact QNMs of
the scalar perturbations in the background of $z=4$ Topological black hole in $4+1$ dimensional Horava-Lifshitz gravity and we study the stability of
this black hole from QNMs. Then, in section III, we discuss our results and conclude.

\section{Quasinormal Modes}
The behavior of the scalar field in the background of $z=4$ Topological black hole in $4+1$ dimensional Horava-Lifshitz gravity is given by the Klein-Gordon equation
 \begin{equation}  \label{KG}
\Box \psi = \frac{1}{\sqrt{-g}}\partial_{\mu}\left(\sqrt{-g}
g^{\mu\nu}\partial_{\nu}\right)\psi=m^2\psi~,
\end{equation}
where $m$ is the mass of the scalar field, $\psi$. In order to obtain exact QNMs, we consider the change of
variables $v=1-\frac{3p}{\Lambda_{\omega}r^2}$, where $p=k\pm \sqrt{c_0}$. So, the metric Eq.~(\ref{metric}) 
can be written as 
\begin{equation}  \label{metric2}
ds^2=\frac{pvc^2}{1-v}dt^2-\frac{3}{4\Lambda_{\omega}v\left(1-v\right)^2}%
dv^2+\frac{3p}{\Lambda_{\omega}\left(1-v\right)}d\Omega_{k}^2~.
\end{equation}
Then, we adopt the ansatz $\psi=R(y)Y(\sum)e^{-i\omega t}$, where $Y$ is a
normalizable harmonic function on $\sum_{D-1}$ which satisfies $\nabla^2Y=-QY
$, with $\nabla^2$ the Laplace operator on $\sum_{D-1}$ and 
\begin{equation}
Q=\left(\frac{d-3}{2}\right)^2+\xi^2~,
\end{equation}
are the eigenvalues
for the hyperbolic manifold. Thus, Eq.~(\ref{KG}) can be written as
\begin{equation}  \label{radial1}
v\left(1-v\right)\partial_{v}^2R(v)+\partial_{v}R(v)+ \left(\frac{3\omega^2}{%
4pvc^2\Lambda_{\omega}}+\frac{Q}{4p}+\frac{3m^2}{4\Lambda_{\omega}\left(1-v%
\right)}\right)R(v)=0~.
\end{equation}
Which, can be written as a hypergeometric equation for K
\begin{equation}  \label{hypergeometric}
\ v(1-v)K^{\prime \prime }(v)+\left[c-(1+a+b)v\right]K^{\prime }(v)-ab
K(v)=0~,
\end{equation}
where the coefficients are given by
\begin{equation}  \label{a}
\ a=-\frac{1}{2}+\alpha+\beta\pm\frac{1}{2p}\sqrt{p\left(p+Q\right)}~,
\end{equation}
\begin{equation}
b=-\frac{1}{2}+\alpha+\beta\mp\frac{1}{2p}\sqrt{p\left(p+Q\right)}~,
\end{equation}
\begin{equation}
c=2\alpha+1~,
\end{equation}
if we perform the decomposition $R(v)=v^\alpha(1-v)^\beta K(v)$, with
\begin{equation}
\alpha_{\pm}= \pm \frac{i\omega}{2c}\sqrt{\frac{3}{p\Lambda_{\omega}}}~,
\end{equation}
\begin{equation}  \label{omega2}
\ \beta_{\pm}=1\pm\sqrt{1-\frac{3m^2}{4\Lambda_{\omega}}}~,
\end{equation}
The general solution of Eq.~(\ref{hypergeometric}) takes the form
\begin{equation}
K=D_{1}F_{1}(a,b,c;v)+D_2v^{1-c}F_{1}(a-c+1,b-c+1,2-c;v)~,
\end{equation}
which has three regular singular points at $v=0$, $v=1$ and $v=\infty$. Here,
$F_{1}(a,b,c;v)$ is a hypergeometric function and $D_{1}$, $D_{2}$ are
constants. Then, the solution for the radial function $R(v)$ is
\begin{equation}  \label{RV}
\ R(y)=D_{1}v^\alpha(1-v)^\beta F_{1}(a,b,c;v)+D_2v^{-\alpha}(1-v)^\beta
F_{1}(a-c+1,b-c+1,2-c;v)~.
\end{equation}

According to our change of variables, when $r\rightarrow
r_{+}$, then $v\rightarrow 0$ and when $r\rightarrow \infty$, then $v\rightarrow 1$. In the vicinity of the horizon, the function $R(v)$ behaves as
\begin{equation}
R(y)=D_1 e^{\alpha \ln v}+D_2 e^{-\alpha \ln v}~,
\end{equation}
where we have used the
property $F(a,b,c,0)=1$. Thus, the scalar field $\varphi$ can be written in the following way
\begin{equation}
\psi\sim D_1 e^{-i\omega(t+ \chi\ln v)}+D_2 e^{-i\omega(t- \chi\ln v)}~,
\end{equation}
where, we have chosen, without loss of generality, the negative signs for $\alpha$,   and
\begin{equation}
\chi=\frac{1}{2c}\sqrt{\frac{3}{\Lambda_{\omega} p}}~.
\end{equation}

The first term represents an ingoing wave and the second one an outgoing
wave on the black hole background. For computing the QNMs, we have to impose that there exist only ingoing waves on the horizon.
This fixes $D_2=0$. Then the radial solution becomes
\begin{equation}  \label{horizonsolutiond}
R(y)=D_1 e^{\alpha \ln v}(1-v)^\beta F_{1}(a,b,c;v)=D_1 e^{-i\omega \chi \ln
v}(1-v)^\beta F_{1}(a,b,c;v)~.
\end{equation}
At infinity ($v=1$), the radial function reads
\begin{eqnarray}  \label{R}
\ R(v) &=& D_1 e^{-i\omega\chi \ln v}(1-v)^\beta\frac{\Gamma(c)\Gamma(c-a-b)%
}{\Gamma(c-a)\Gamma(c-b)} F_1(a,b,a+b-c,1-v)  \notag \\
&& +D_1 e^{-i\omega\chi\ln v}(1-v)^{c-a-b+\beta}\frac{\Gamma(c)\Gamma(a+b-c)%
}{\Gamma(a)\Gamma(b)}F_1(c-a,c-b,c-a-b+1,1-v)~,
\end{eqnarray}
we have applied in Eq. (\ref{horizonsolutiond}) the Kummer's formula for the
hypergeometric function \cite{M. Abramowitz},
\begin{equation}
F_{1}(a,b,c;v)=\frac{\Gamma(c)\Gamma(c-a-b)}{\Gamma(c-a)\Gamma(c-b)}
F_1(a,b,a+b-c,1-v)+(1-v)^{c-a-b}\frac{\Gamma(c)\Gamma(a+b-c)}{%
\Gamma(a)\Gamma(b)}F_1(c-a,c-b,c-a-b+1,1-v)~.
\end{equation}

For, $\beta_+>2$ the field at infinity is regular if the gamma function $%
\Gamma(x)$ has the poles at $x=-n$ for $n=0,1,2,...$. Which, yields 
\begin{equation}\label{1w1}
\omega_1 =ic\sqrt{\frac{\Lambda_{\omega}}{3p}}\left[ \left(1+\sqrt{4-\frac{%
3m^2}{\Lambda_{\omega}}}+2n\right)p+\sqrt{p\left(p+Q\right)} \right]~.
\end{equation}

Let us now investigate the stability of $z=4$ Topological black hole in $4+1$ dimensional Horava-Lifshitz gravity, for the positive branch, $p=-1+\sqrt{c_0}$, $z=4$ Topological black hole in $4+1$ dimensional Horava-Lifshitz gravity is stable due to the imaginary part of the QNMs is negative, because $p<0$. Then, for the negative branch, $p=-1-\sqrt{c_0}$. There are two cases:
\begin{itemize}
\item
$p+Q>0$. The QNMs are given by Eq. (\ref{1w1}) with $p=-1-\sqrt{c_0}$, which can be written as
\begin{equation}\label{w1}
\omega_1 =-c\sqrt{\frac{-\Lambda_{\omega}(p+Q)}{3}}+ic\sqrt{\frac{\Lambda_{\omega}}{3p}}\left(1+\sqrt{4-\frac{%
3m^2}{\Lambda_{\omega}}}+2n\right)p ~.
\end{equation}
Therefore, $z=4$ Topological Horava-Lifshitz black hole in $4+1$ dimensions is stable.
\item
$p+Q<0$. The QNMs, given by Eq. (\ref{w1}) can be written as
\begin{equation}
\omega_1 =-ic\sqrt{\frac{\Lambda_{\omega}p}{3}}\left(1+\sqrt{4-\frac{%
3m^2}{\Lambda_{\omega}}}+2n-\sqrt{1+\frac{Q}{p}} \right)~.
\end{equation}
which is purely imaginary and the stability is guaranteed due to $1+\frac{Q}{p}<1$.
\end{itemize}

Such as we have showed, the quasinormal modes for scalar perturbations                                                                                                                                                                                                                                                                                                                                                                                                                                                                                                                                                                                                                                                                                                                                                                                                                                                                                                                                                                                                                                                                                                                                                                                                                                                                                                                                                                                                                                                                                                                                                                                                                                                                                                       can be found by imposing the vanishing Dirichlet boundary condition at infinity. Adittionally, we can consider that the flux vanishes at infinity or vanishing Neumann boundary condition at infinity. It is due to that in
asymptotically anti-de Sitter, a negative mass square for a scalar field is consistent, in agree with the analogue to the Breitenlohner-Freedman condition that any effective mass must satisfy in order to have a stable propagation, \cite{Breitenlohner:1982bm, Breitenlohner:1982jf}. That is, Dirichlet boundary condition leads to the same quasinormal modes for $m^2 > 0$ but does not lead to any quasinormal modes for $m^2 < 0$. So, quasinormal modes can be obtained by considering vanishing Neumann boundary conditions and in the context of conformal field theory, due to the correspondence AdS/CFT \cite{Maldacena:1997re}, the quasinormal modes match with the dual operators, \cite{Birmingham:2001pj}. So, in order to consider Neumann boundary condition at infinity, $v\rightarrow 1$, the flux
\begin{equation}
F=\frac{\sqrt{-g}g^{rr}}{2i}\left( R^{\ast }\partial _{r}R-R\partial
_{r}R^{\ast }\right)~,
\end{equation}
at infinity, is given by
\begin{eqnarray}
F\left( v\rightarrow 1\right)  &=&-\frac{6\gamma p^{2}}{\Lambda _{\omega }}%
\left\vert D_{1}\right\vert ^{2} Im (2\left( 1-\beta \right)
B_{1}^{\ast }B_{2}+\left( -i\omega \chi -\frac{ab}{a+b-c}\right) \left\vert
B_{1}\right\vert ^{2}\left( 1-v\right) ^{2\beta -1}~, \\
&&+\left( -i\omega \chi -\frac{\left( c-a\right) \left( c-b\right) }{c-a-b+1}%
\right) \left\vert B_{2}\right\vert ^{2}\left( 1-v\right) ^{3-2\beta })~,
\end{eqnarray}
where,
\begin{equation}
B_{1}=\frac{\Gamma \left( c\right) \Gamma \left( c-a-b\right) }{\Gamma
\left( c-a\right) \Gamma \left( c-b\right) }~,
\end{equation}
\begin{equation}
B_{2}=\frac{\Gamma \left( c\right) \Gamma \left( a+b-c\right) }{\Gamma
\left( a\right) \Gamma \left( b\right) }~.
\end{equation}
For, $\frac{4}{3}\Lambda _{\omega }<m^{2}<0$,
$1<\beta _{+}<2$ and $0<\beta _{-}<1$. So, if $B_{1}=0$, the flux vanishes at infinity
if $3-2\beta >0$.  For, $\beta =\beta _{+}$
is possible to obtain a new set of quasinormal modes.
Note that, the condition $3-2\beta _{+}>0$ implies that
$\sqrt{-\Lambda _{\omega }}<Im\left(m\right) < \sqrt{-\frac{4}{3}\Lambda _{\omega }}$. Therefore, the QNMs are given by
\begin{equation}\label{2w2}
\omega_2 =ic\sqrt{\frac{\Lambda _{\omega }}{3p}}\left[ \left( 1-\sqrt{4-\frac{%
3m^{2}}{\Lambda _{\omega }}}+2n\right) p-\sqrt{p\left( p+Q\right) }\right] ~,
\end{equation}
where, we have considered $c-a|_{\beta _{+}}=-n$ o $c-b|_{\beta _{+}}=-n$. The condition $B_{2}=0$ leads to the same quasinormal modes that we have found by imposing Dirichlet boundary condition.

Let us now investigate the stability of $z=4$ Topological black hole in $4+1$ dimensional Horava-Lifshitz gravity and let us consider the positive branch, $p=-1+\sqrt{c_0}$. In this case, $z=4$ Topological black hole in $4+1$ dimensional Horava-Lifshitz gravity is stable due to the imaginary part is negative, in the range $\sqrt{-\Lambda _{\omega }}<Im\left(
m\right) <\sqrt{-\frac{4}{3}\Lambda _{\omega }}$. On the hand, for the negative branch, $p=-1-\sqrt{c_0}$. There are two cases:
\begin{itemize}
\item
$p+Q>0$. The QNMs are given by Eq. (\ref{2w2}) with $p=-1-\sqrt{c_0}$, that is
\begin{equation}\label{w2}
\omega_2 =c\sqrt{\frac{-\Lambda_{\omega}(p+Q)}{3}}+ic\sqrt{\frac{\Lambda_{\omega}}{3p}}\left(1-\sqrt{4-\frac{3m^2}{\Lambda_{\omega}}}+2n\right)p ~.
\end{equation}
Therefore, $z=4$ Topological black hole in $4+1$ dimensional Horava-Lifshitz gravity is stable.
\item
$p+Q<0$. The QNMs, given by Eq. (\ref{w2}) can be written as
\begin{equation}
\omega_2 =-ic\sqrt{\frac{\Lambda_{\omega}p}{3}}\left(1-\sqrt{4-\frac{3m^2}{\Lambda_{\omega}}}+2n+\sqrt{1+\frac{Q}{p}} \right)~.
\end{equation}
which is purely imaginary and the stability of $z=4$ Topological black hole in $4+1$ dimensional Horava-Lifshitz gravity is guaranteed in the range $\sqrt{-\Lambda _{\omega }}<Im\left(
m\right) <\sqrt{-\frac{4}{3}\Lambda _{\omega }}$.
\end{itemize}

\section{Conclusions}

In this work we have studied scalar perturbations in the background of a $z=4$ Topological black hole in $4+1$ dimensional Horava-Lifshitz gravity and we have found the quasinormal modes using the Dirichlet and Neumann boundary conditions. The QNMs that we have found, generally are complex but for a certain range of parameters they are purely imaginary. Then, from these QNMs we
have analyzed the stability of this black hole and we have found that  $z=4$ Topological black hole in $4+1$ dimensional Horava-Lifshitz gravity is stable and the relaxation time $\tau$ for a thermal state to reach thermal equilibrium in the boundary conformal field theory is finite, according to the AdS/CFT correspondence. Is interesting note that from a thermodynamics point of view the $z=4$ Topological black hole in $4+1$ dimensional Horava-Lifshitz gravity is stable \cite{Cai:2009ar}, which agree with the QNMs point of view.

\section*{Acknowledgments}

Y. V. is supported by FONDECYT grant 11121148, and by Direcci\'{o}n de Investigaci\'{o}n y Desarrollo, Universidad de La Frontera, DIUFRO DI11-0071.


\appendix

\end{document}